# Readout method based on PCIe over optical fiber for CBM-TOF super module quality evaluation

Jianhui Yuan, Ping Cao, Xiru Huang, Chao Li, Qi An

*Abstract*—The Compressed Baryonic Matter (CBM) experiment will investigate the quantum chromodynamics (QCD) phase diagram at high net baryon densities and moderate temperatures. CBM Time of Flight (TOF) system is composed of super modules containing high performance Multi-gap Resistive Plate Chambers (MRPCs). During the mass production, each super module assembled with MRPCs needs quality evaluation, which includes time measurement and data readout. Read out electronics encounter the challenge of reading data from a super module at a speed of about 6 Gbps. In this paper, a read out method based on Peripheral Component Interconnect Express (PCIe) over optical fiber is proposed for CBM-TOF super module quality evaluation. The digitized data from super module will be concentrated at the front-end electronics, and then be transmitted to a PCIe switch module (PSM) over optical fiber using PCIe protocol. The PSM is directly plugged into the motherboard via gold fingers at the backend data acquisition server. With this readout method, a high-speed transmission rate can be reached. Furthermore, a PSM can receives data from several super modules simultaneously, which is important to improve the evaluation efficiency. This readout method simplifies the architecture of readout electronics and supports long distance transmission between frontend and backend.

## I. INTRODUCTION

THE Compressed Baryonic Matter (CBM) is one of the core projects of the future Facility for Antiproton and Ion Research (FAIR) in Darmstadt. Time of Flight (TOF) system, one of the detectors of the CBM experiment, is designed to identify hadrons. The CBM-TOF is composed of 6 kinds of super modules (M1~M6) containing high performance Multi-gap Resistive Plate Chambers (MRPCs) [1], as is shown in Fig. 1. During the mass production, the quality evaluation of super modules encounters the challenge of reading data from a super module at a speed of about 6 Gbps [2]. Traditionally in high-energy experiments, data from FEE are collected in readout modules [3, 4, 5], which are distributed in Versa Module Eurocard (VME) or Peripheral Component Interconnect extensions for Instrumentation (PXI) crates. And each module transmits data to the crate controller through VME or PXI backplane. At end, the crate controller sends data to data acquisition (DAQ) system by Ethernet. However, limited by the bandwidth of the crate backplane, VME or PXI crate is not able to meet the requirement of about 6 Gbps data rate for a CBM-TOF super module.

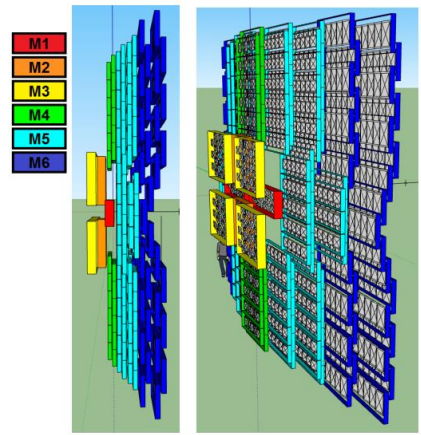

Fig. 1. Side view of CBM-TOF wall with super modules.

In this paper, a readout method based on Peripheral Component Interconnect Express (PCIe) over optical fiber is proposed for CBM-TOF super module quality evaluation.

## II. ARCHITECTURE OF READOUT ELECTRONICS

### A. Front-end electronics (FEE)

During the mass production, for the quality control of super module containing with MRPCs, a high density Time-to-Digital Converter (TDC) prototype with sandwich structure is proposed by J. Zheng et al [6]. The FEE contains a Time Over Threshold (TOT) Feeding Board (TFB), ten TDC cards, a TDC Readout Motherboard (TRM) and a Small Form-factor Pluggable (SFP) daughterboard. The structure of FEE is illustrated in Fig. 2. Signals from MRPCs will be converted into digital signals, and be output to FEE.

This work is supported by National Basic Research Program (973 Program) No. 2015CB85906.
The authors are with the State Key Laboratory of Particle Detection and Electronics, University of Science and Technology of China, Hefei 230026, China (email: cping@ustc.edu.cn).

J. Yuan and P. Cao are with School of Nuclear Science and Technology, University of Science and Technology of China, Hefei 230026, China.
X. Huang, C. Li, and Q. An are with the Department of Modern Physics, University of Science and Technology of China, Hefei 230026, China.

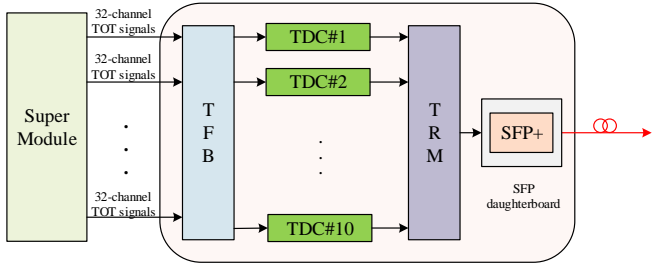

Fig. 2. The structure of FEE.

The TFB transfers 320-channel TOT signals from a super module to ten TDCs. And the signals are distributed equally between ten TDCs. Each of the ten TDCs receives 32-channel TOT signals. Data of time information that TDC resolves from TOT signals will be transmitted to TRM. The TRM is responsible for collecting data from ten TDCs and sending them to SFP daughterboard. The SFP daughterboard transmits data to back-end electronics via optical links.

### B. Back-end readout electronics

To transmit data at such a high rate, a readout method based on PCIe over optical fiber is proposed in this paper. Fig. 3 explained the structure of this readout method.

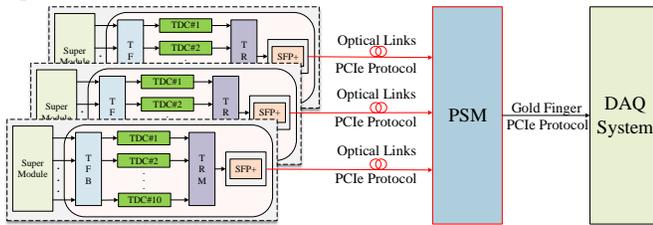

Fig. 3. The readout scheme for super module evaluation.

A PCIe interface is implemented in the Field Programmable Gate Array (FPGA) on each TRM. The TRM utilizes this PCIe interface to transmit data over optical fiber to the back-end electronics. The PSM receives the data from TRM with SFP plus transceivers. The PSM is designed as a PCIe add-in card, so it is convenient to plug a PSM into the motherboard of a DAQ server. The PSM works like a transparent bridge of PCIe tree. It routes data from different TRMs, each of which corresponds to a super module, to the DAQ system through gold finger. And each TRM seems like a PCIe device to the DAQ server.

Traditional cables limit the distance between front-end electronics and back-end electronics because that long distance transmission causes high-speed signal attenuation. The deployment of optical links in this method allows long distance transmission and crates an electronic isolation between front-end electronics and back-end electronics.

### III. THE IMPLEMENTATION OF PSM

PSM is the most important part of this readout method. It is responsible for offering a high bandwidth between the FEE and the DAQ system. And it works as a PCIe switch that connects multiple TRMs with a server of the DAQ system.

As shown in Fig. 4, a PCIe system interconnect switch chip is used for PCIe packet switching. The PSM receives data from TRMs via optical fiber with SFP+ transceivers. To minimize the signal degradation effects such as crosstalk and inter-symbol interference, PCIe repeaters are used in lanes between SFP+ transceivers and the PCIe system interconnect switch chip. As shown in Fig. 5, the PSM is designed as a PCIe add-in card, so PCIe packets could be transferred to the motherboard of the DAQ server via gold fingers. Extra PCIe lanes are provided with SFP+ for possible optical links from PSM to the DAQ system. Configuration file that the PCIe system interconnect switch chip need is provided by an Electrically Erasable Programmable Read-Only Memory (EEPROM). The PCIe switch chip can also be configured by software runs in the DAQ system. And a PCIe socket is provided for the convenient of test.

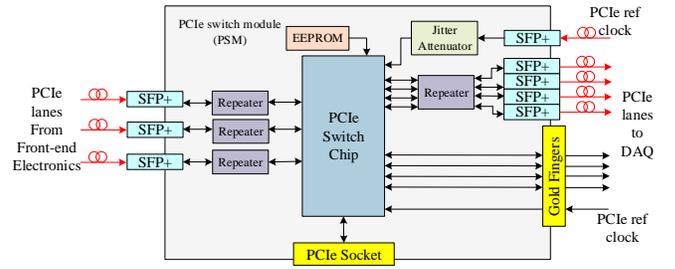

Fig. 4. The block diagram of PSM.

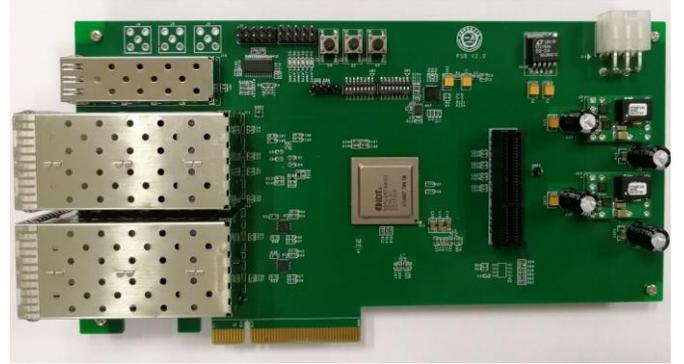

Fig. 5. PCIe Switch Module.

The switch chip could be configured by EEPROM or by software through upstream PCIe lane. Connected with RC, port 0 and port 8 are set as upstream port with Peer-to-Peer Bridge (P2P) and Direct Memory Access (DMA) function. The other ports are set as downstream port and are connected with Endpoint (EP). Ports could be configured to a single partition or multiple partitions.

### IV. CONCLUSION

The super module detector of CBM-TOF experiment is provided with high density of electronics channels. And that brings the necessity for readout electronics with high speed ability of data readout. Limited by the bandwidth of the crate backplane, traditional VME or PXI technique could not meet the requirement. In this paper, a readout method based on PCIe over optical fiber is proposed for CBM-TOF super module quality evaluation. With this readout method, the digitized time data from FEE are sent to a readout module, which named with PSM, via optical fiber with PCIe protocol. The PSM plays the role of a switch between the FEE and the DAQ servers.

Furthermore, besides the application of CBM-TOF super module quality evaluation, this readout method could be easily applied to other high energy physical experiments that require high speed data readout ability.